\documentstyle{mn}
\title[Fourier-phase analysis\ldots]{Fourier-phase analysis of the
orbiting bright-spot model for AGN variability}
\author[V. Karas]{V. Karas\thanks{Also at Scuola Internazionale Superiore
 di Studi Avanzati, Trieste;
 Department of Astronomy and Astrophysics, G\"oteborg University and
 Chalmers University of Technology, G\"oteborg}\\
 Astronomical Institute, Charles University, \v{S}v\'edsk\'a 8,
 CZ-150 00 Prague, Czech Republic}
\date{Received }
\pagerange{\pageref{firstpage}--\pageref{lastpage}}

\newcommand{\beq}{\begin{equation}}
\newcommand{\eeq}{\end{equation}}

\newcommand{\etal}{{\rm et~al.}}
\newcommand{\dg}{^{\rm o}}
\newcommand{\sfrac}[2]{{\textstyle\frac{#1}{#2}\,}}
\def\spose#1{\hbox to 0pt{#1\hss}} \def\lta{\mathrel{\spose{\lower
3pt\hbox{$\mathchar"218$}} \raise 2.0pt\hbox{$\mathchar"13C$}}}
\def\gta{\mathrel{\spose{\lower 3pt\hbox{$\mathchar"218$}} \raise
2.0pt\hbox{$\mathchar"13E$}}}
\begin{document}
\label{firstpage}

\maketitle

\begin{abstract}
Fourier power spectra and phases of a signal from a large number
of radiating sources orbiting around a black hole are investigated.
It is assumed that the individual sources (bright spots) are located
in an accretion disc and their lifetime exceeds the corresponding
orbital period. This model is relevant for the short-time X-ray
variability of active galactic nuclei. Previous works on this subject
were mostly concentrated on temporal characteristics and power spectra
of observed light curves. In our present contribution, Fourier phases
are brought into consideration and studied systematically for a broad
range of input parameters. In particular, conditions for the
phase coherence are discussed. It is shown that one can discriminate between
the two classes of models which are currently under consideration---orbital
motion of a large number of sources versus short-lived independent
flares---although parameters of the model are not completely arbitrary.
It is also shown that
predicted power spectra depend rather strongly on the spot
distribution across the disk surface. We conclude that the orbital motion
of the spots cannot be the only reason for the source fluctuations,
but it certainly influences observational properties of the source intrinsic
variability.
\end{abstract}



\section{Introduction}
Observational effort of recent decades brought
an extensive amount of data about the variability properties of
active galactic nuclei (AGNs) in the range of wavelength from
radio to hard X-rays (Duschl, Wagner \& Camenzind 1991; Miller \& Wiita 1991).
It is widely presumed, although not yet
proven, that massive black holes with
accretion discs reside in cores of AGNs (Blandford, Netzer \& Woltjer 1990).
Within this model which is also accepted in the present contribution,
most of the X-rays originate inside inner regions of AGNs,
a few or a few tens of
gravitational radii from the center. The observed signal shows irregular,
featureless fluctuations with a very complex
behaviour at frequency about $10^{-5}$--$10^{-2}\,$Hz.
The signal has traditionally been analyzed by statistical approaches,
the Fourier analysis in particular (Feigelson \& Babu 1992).
In the above-mentioned restricted range of frequency, power-spectrum density
of the fluctuating signal can be
represented by a power-law: $P(\omega)\propto\omega^{-\alpha}$
with $1\lta\alpha\lta2$ (Lawrence \etal{} 1987; McHardy \& Czerny 1987;
Mushotzky, Done \& Pounds 1993; Lawrence \& Papadakis 1993).
This form of the power spectrum can
be accommodated by several physical models of the
central region which have been proposed in the literature, restricting
partially the parameter space of each of them. However,
the power-spectrum itself does not allow us to discriminate between
intrinsically different models or exclude some of the proposed scenarios
from further consideration. Models of the AGN X-ray variability
refer to various instabilities in the inner parts of
the accretion disc, the disc corona or the jet which probably operate in
a nonlinear regime (Krolik \etal{} 1991; Mineshige, Ouchi \&
Nishimori 1994; Ipser 1994; Kanetake, Takeuti \& Fukue 1995).
Another viable approach considers the observed variability as a direct
consequence of the orbital motion of numerous irregularities
(regions of enhanced emissivity, or bright spots) on the surface
of the accretion disc (Abramowicz, Bao, Lanza \& Zhang 1991;
Wiita, Miller, Carini \& Rosen 1991). The latter scenario has been
formulated on a phenomenological level. The origin of the spots
remains an open question and the mechanism, by its definition,
turns out to be relevant only if the lifetime of individual irregularities
is longer than the local orbital period. In this respect, the possible
existence of vortices (Abramowicz, Lanza, Spiegel \& Szuszkiewicz 1992) and
spiral waves (Chakrabarti \& Wiita 1994) in accretion discs has been
discussed in the literature. The bright-spot model has been also applied to
to the UV microvariability of AGNs (Mangalam \& Wiita 1993).

In the present contribution, we wish to clarify
whether there is enough useful information in the Fourier spectra
to indicate that the observed fluctuations are due to orbital motion of
separate (mutually unrelated) sources of radiation rather than virtually
random flares. It has been stressed repeatedly that an appropriate approach
to data analysis should take the physical nature of the system into
account (cf.\ Vio \etal{} for references). It appears conceivable,
however, that one can discriminate between the two alternatives---orbital
motion of sources each of which survives many orbital periods versus
short-lived flares---even when parameters of the model are not fully
determined. The reason for this hope is apparent: orbital motion
of the source modifies the intensity of observed radiation periodically.
The relative change of the radiation flux (count rate)
is particularly large when the disc is seen edge-on. This effect
is mainly due to gravitational focusing (Bao 1992; Karas, Vokrouhlick\'y
\& Polnarev 1992, and references cited therein), and it turns out that
light rays must be calculated properly by solving the geodesic equation
when inclination of the observer is greater than about $70\dg;$ otherwise,
simulated light-curve profiles are wrong.
On the other hand, the hope of finding a signature of the orbital motion in
fluctuating light curves may be false when the sources
are spread through a large range of radii from the black hole, so that
their orbital periods differ too much. Also, the harmonic
content of the light-curve profiles is then large.
These facts are well known from
studies of the power-spectra (McHardy 1989; Zhang \& Bao 1991) and indicate
that the bright-spot model should be considered as one of
viable explanations of the quasi-periodic oscillations (see, e.g.,
Miller \& Park [1995] for a recent list of references).

In the next section, properties of the Fourier transform relevant
for the present work are briefly summarized. Fourier phases are discussed
because it turns out that power spectra do not contain enough information.
Details of our model of orbiting sources are then specified and
results of a systematic exploration of its parameter space are illustrated.

\section{Fourier power spectra and phases}
Power-spectrum analysis is useful for detecting periodicities present in
a signal, and for this reason it has been discussed extensively in the
astronomical literature and applied to various problems.
The power spectrum, however, contains less
information than original data or their Fourier transform from
which the power spectrum has been constructed.
The missing portion of information
can be reconstructed from the Fourier phase. Physical interpretation of
the Fourier phase and
its relation to properties of the corresponding system is much
less understood than the power spectrum.
Analysis of simple model cases suggests
that it may be possible to reveal hidden periodicities by studying Fourier
phases. Lanczos \& Gellai (1975) performed the Fourier analysis
of random, computer-generated sequences and noticed patterns in
the distribution of the Fourier phases. Their aim was to investigate
periodicities which are present in
pseudo-random numbers. (A historical note which is added to
the report by Lanczos \& Gellai suggests that the
original motivation to consider
Fourier analysis of random sequences stemmed from Einstein's [1915]
article.) Non-uniform distribution of phases and their clustering
to certain angles is usually referred to as a phase coherence.
Krolik, Done \& Madejski (1993) have explored the phase coherence of X-ray
light curves of five Seyfert galaxies observed by EXOSAT but found
no statistically significant coherence in their data.
(It has been demonstrated later that the EXOSAT
data for one of their sources, then a popular candidate for the
AGN periodic variability, were dominated by a cataclysmic variable
in the Galaxy; cf.\ Madejski, Done, Turner, Mushotzky et al.\ 1994.)

\subsection{Details of the model}

\begin{figure}
 \vspace{8.5cm}
\caption{Fourier analysis of the signal from a single orbiting spot is
presented as an illustrative example which helps to understand
more complicated
situations with a large number of spots. No noise component was
added to the signal. Parameters of this figure are summarized in the
first row of Tab.~\protect\ref{tab1}. The counting rate (a) is
mean subtracted and normalized. Notice how the phase coherence of the
signal is reflected in panels (d)--(f).}
\label{fig1}
\end{figure}

In this paragraph we specify the bright-spot model with sufficient
generality and ask whether its relevance for understanding AGN variability
can be tested by supplementing discussion of power spectra with analysis
of Fourier phases. The mass of the central black hole in AGNs is usually
estimated to be in the range of $M\approx10^6$--$10^{11}\,M_\odot.$
The mass of the accretion disc is neglected and gravitational field is
described by a vacuum spacetime of a nonrotating (Schwarzschild)
black hole (Misner, Thorne \& Wheeler 1973).
The corresponding gravitational radius of a black hole is
$R_g=2c^{-2}GM\approx10^{-5}M_8\,$pc, and characteristic
time $t_g=2c^{-3}GM\approx10^3M_8\,$sec, where $M_8\equiv M/(10^8M_\odot).$
Geometrized units, $c=G=1,$ are used hereafter.
All lengths and times are made dimensionless by expressing them
in units of $M.$

It is assumed that the disc is geometrically thin and rotates with the
Keplerian circular frequency. A generalization to geometrically thick
discs is straightforward once the disc height and its angular velocity
are specified as functions of radius. The main difference between
geometrically thick and geometrically thin discs consists in a
possibility of eclipses when observer inclination is large, but we
neglect these additional free parameters of the model in the present
contribution; it turns out that the restriction is irrelevant to the
question whether resulting light curves are phase coherent or not.
Profiles of frequency-integrated light curves of an orbiting source of
light were originally studied by Cunningham \& Bardeen (1973). Later, a
number of authors employed various computational schemes which are
needed for efficient evaluation of observed radiation flux resulting as
a superposition of contributions from a large number of individual
sources. In the present work, an analytical fitting-formula for the
observed counting rate $F(\varphi,R_s,\theta_0)$ from each of the spots
(in arbitrary units) was employed:
\begin{eqnarray}
F(\varphi,R_s,\theta_0) & = & \left(\frac{p_3\cos\theta_0}{R_s}+
  p_7\left(R_s-1\right)^{2/5}\right)\cos^{-2/3}\theta_0 \nonumber \\
 & \times &
  \Bigg[1+\sin\left[2\pi(\varphi+p_4\,R_s-\sfrac{3}{5})+
  \sfrac{\pi}{2}\right]\Bigg]^{z_1}    \nonumber \\
 & + &
  \cos^{-2}\theta_0\left(p_1+p_6R_s^{1/3}\right)\,\exp\left[-p_2z_2\right],
 \label{fit}
\end{eqnarray}
\[z_1\equiv p_5\cos^{1/2}\theta_0+p_8\cos^{3/2}\theta_0,\]
\[z_2\equiv\left|\varphi-\sfrac{1}{2}\right|^{9/5};\]
$R_s$ is the radius of the orbit of the source, $\theta_0$ is observer
inclination $(\theta_0=90\dg$ corresponds to the edge-on view of the
disk), and $\varphi\propto tR_s^{-3/2}$ is the normalized phase of the
periodic signal $(0\leq\varphi\leq1)$. Eq.~(\ref{fit}) takes into
account the fact that observed energy of radiation is different from
radiation energy in the local rest frame of the source of light (Doppler
effect, gravitational redshift). Focusing of light by the central black
hole is also included.

\begin{figure}
\vspace{7.8cm}
\caption{As in Fig.~\protect\ref{fig1} but with 100 spots orbiting in the
range of radii $6\,R_g\leq R_s\leq25\,R_g.$ Low values of $\sigma$ and $p$
(Tab.~\protect\ref{tab1}) indicate the phase-coherence which can also be
inferred from varying histograms in the panel (d),
although the effect is less apparent than in Fig.~\protect\ref{fig1}.}
\label{fig2}
\end{figure}

Several restrictions have been imposed in derivation of the
fitting-formula (\ref{fit}) (Karas 1996): (i)~The bright spot is located
in a Keplerian circular orbit in the equatorial plane of a black hole;
(ii)~Radius of the orbit satisfies condition $3\,R_g\leq{}R_s\leq45\,R_g$
(iii)~Local emissivity of the bright spot is isotropic (in the rest frame
comoving with the spot) and it decreases exponentially from the center
of the spot; (iv)~Characteristic size $d$ of the spot (defined as a
distance from the spot center to the point where emissivity decreases by
factor $1/e$ with respect to its value in the spot center) satisfies
$d\ll R_s$; (v)~Inclination angle satisfies condition
$20\dg\leq\theta_0\leq 80\dg$.

The numerical code which produced simulated light curves was described
by Karas, Vokrouhlick\'y \& Polnarev (1992), while derivation and further
details concerning eq.~(\ref{fit}) have been discussed in the
above-mentioned paper (Karas 1996).  Here, we only give the values of
parameters $p_k$ which were obtained by non-linear least-square fitting
to numerically simulated light curves: $p_1=0.021696$, $p_2=190.7236$,
$p_3=0.3476$, $p_4=-0.0018$, $p_5=3.5106$, $p_6=-3.6\times10^{-5}$,
$p_7=0.0124$, $p_8=-0.0231.$ Only a non-rotating black hole is
considered in the present contribution because parameters $p_k$ are not
sensitive to the black-hole angular momentum. It is worth noting that
the analytical form of eq.~(\ref{fit}) proves to be more efficient than
previous numerical approaches, while preserving the characteristic shape
of the light curve to which gravitational focusing and the Doppler
effect both contribute. We found no indication that our results are
affected by the above-given assumptions (i)--(v) which were needed only
in derivation of the fitting-formula (\ref{fit}) from simulated data
by the least-square procedure.

\begin{figure}
 \vspace{7.8cm}
\caption{As in Fig.~\protect\ref{fig2} but with $5\,$\% noise component added.
One observes a decrease in the phase coherence due to the noise. See the
text for a detailed description.}
\label{fig3}
\end{figure}

Individual contributions to the total flux were generated with random
orbital phases and the resulting flux was submitted to the Fourier
analysis. The light-curve profile from a single source (and without a
noise) is, naturally, periodic in time. The idea of these models is that
a superposition of many contributions with different periods results in
a fluctuating signal. According to the present status of the bright-spot
model, it appears that the fluctuations due to the orbital motion cannot
be {\it the only reason\/} of AGNs variability but they are certainly
coupled with intrinsic fluctuations and influence the resulting
observational properties of AGN. Denoting the total flux from $N_s$
sources by
\beq
F(t,\theta_0)=\sum_{n=1}^{N_s}F_n(t,R_s,\theta_0),
\eeq
and carrying out the Fourier transform, $F(t)\rightarrow\hat{F}(\omega)$,
one obtains for the Fourier power spectrum and the Fourier phase
(Born \& Wolf 1964):
\beq
P(\omega)=\sfrac{1}{2}\lim_{t\rightarrow\infty}\left[t^{-1}
 \left|\hat{F}(\omega)\right|^2\right],
\eeq
\beq
\phi(\omega)=\arctan\frac{{\Im}m[\hat{F}(\omega)]}{{\Re}e[\hat{F}(\omega)]}\,.
\eeq
The phase $\phi$ takes values from interval $\langle-\pi,\pi\rangle.$
The two boundary values, $\pm\pi,$ are identified with each other,
and play thus no particular role.
It is then possible to smooth the power spectrum and, within a restricted
range of frequency, to approximate the resulting curve by a power-law,
\beq
P(\omega)=\omega^{-\alpha}.
\label{alpha}
\eeq
The phase coherence is expected between harmonics corresponding to the
spots located at the same radius, but it is difficult to estimate how
the coherence will be suppressed when a range of radii is considered. A
large number of model parameters can conveniently be investigated by
checking whether $\alpha$ is within the expected interval of values and
then, for selected cases, performing a formal test for the phase
coherence.

We define an average relative distance of $M$ phase points by
\beq
\sigma=\frac{1}{\pi(M-1)}\sum_{m=2}^{M}d_m
\label{sigma}
\eeq
with
\[d_m\equiv\mbox{Min}\left\{\left|\phi_m-\phi_{m-1}-k\pi\right|
 \right\},\quad k=-1,0,1.\]
Frequency is assumed to fall within a limited interval
$\langle\omega_{\min}\equiv1/T_{\max},$
$\omega_{\max}\equiv1/T_{\rm \min}\rangle$ where
$T_{\min}\equiv T(R_{\min})$ and $T_{\max}\equiv T(R_{\max})$
are, respectively,
orbital periods at the minimum and the maximum radius of the spot
distribution.
$M,$ the total number of the phase points in the spectrum,
can, in principle, acquire an arbitrary large value in our tests, but
it is very restricted in real observations because
of their temporal sampling. Apparently, $0\leq\sigma\leq1,$ with values
near zero indicating the kind of coherence where the phase changes only
slowly while values near unity indicate complete incoherence of the
phase points.

Another, also rather crude test of the phase coherence is taken from
\S3.2 of Krolik \etal{} (1993). The frequency-phase plane is binned and
the consistency of the distribution with a uniform distribution is
tested by calculating $\chi^2$ statistic and corresponding probability
$p.$ The binning must be chosen appropriately so that there are enough
phase points in each of the bins, $M_i$. We accepted a uniform binning
into two frequency bins between $\omega_{\min}$ and $\omega_{\max},$
and four bins in phase angles. Thus we write, in standard notation,
\beq
 \chi^2=8M^{-1}\sum_{i=1}^8\left(M_i-\sfrac{1}{8}M\right)^2,\quad
 p=\Gamma\left(\sfrac{7}{2},\sfrac{1}{2}\chi^2\right).
 \label{chi2}
\eeq
Let us note that the two tests described above are not
equivalent in general, but they are both capable of indicating
the phase coherence which occurs, for example, in a simple sine-type signal.
Since no formal test of the phase coherence with a general
validity is available, a direct inspection of graphs showing the phases
also helps.

We studied a simplified model in which individual sources are
identified with bright spots. They are distributed across
the disc surface with the number density and maximum observed fluxes
being given by power-laws:
\beq
 n(R)\propto R^{1-\alpha_n},\quad F_{n,{\max}}\propto R^{-\alpha_i}.
 \label{alphan}
\eeq
In addition, we assumed that the spots have a fixed radius and are
destroyed after several revolutions. We checked that results are not
very sensitive to the latter two assumptions about the size and the
lifetime of the spots; it is only required that the size $d$ satisfies relation
$0<d\ll{}R_s$ and the lifetime $t_s\gg{}R_s^{3/2}.$ However,
power-spectrum indexes are rather sensitive to the spots distribution,
as discussed later. Our work extends
previous calculations in which information in Fourier phases was not taken
into account and
the light-curve profiles of individual spots were either approximated
by a simple box-type pulse (Zhang \& Bao 1991; Abramowicz \etal{} 1991)
or their parameters were assumed to fall in a more restricted range (Bao 1992).

\subsection{Results}
Description of the spots' characteristics in terms of power-laws
represents a simplified model. We carried out a systematic investigation
of the parameter space of this model. Our results are illustrated in
Figures \ref{fig1}--\ref{fig4}. First, it is useful to consider a single
spot because in this case the corresponding light curve has a
particularly simple form. The situation is explored in Figure~\ref{fig1}
which consists of six panels:

\begin{table*}
 {\centering
 \caption{Fourier analysis of the signal generated by bright spots
 orbiting around a Schwarzschild black hole.
 Selected cases are illustrated in
 corresponding Figures, as indicated in the first column of the Table.}
 \label{tab1}
 \begin{tabular}{crrccccccr@{.}llcccr@{.}llc}
\multicolumn{7}{c}{} &
\multicolumn{5}{c}{No noise} &
\multicolumn{1}{c}{} &
\multicolumn{5}{c}{$5\,$\% noise component added} &
\multicolumn{1}{c}{} \\ \cline{8-12} \cline{14-18}
 Fig. & $N_s$ & $R_{\min}$ & $R_{\max}$ & $\theta_0$ & $\alpha_n$
 & $\alpha_i$ & $\alpha$ & $\sigma$ & \multicolumn{2}{c}{$\chi^2$} & $~~p$
 & & $\alpha$ & $\sigma$ & \multicolumn{2}{c}{$\chi^2$}
 & $~~p$ & $\tilde{\alpha}\rule{0mm}{2.5ex}$ \\ \hline \hline
\protect\ref{fig1}
   & 1& 56~~ & 56& 25& -- & -- &  -- & 0.021& 107&8& 0& & -- & 0.388&
    30&55 &0.0050 & -- \\
\protect\ref{fig2}--\protect\ref{fig3}
   & 100&  6~~ & 25& 50& 0& 1& 3.0& 0.485& 662&2& 0& & 2.9& 0.607&
    210&96& 0& 0.3 \\
\protect\ref{fig4}
   & 200&  6~~ & 25& 20& 0& 1& 3.2& 0.537& 5&77& 0.567& & 3.1& 0.640&
    5&01& 0.659& 0.3 \\
-- & 200& 10~~ & 30& 80& 0& 0& 1.8& 0.611& 9&87& 0.196& & 1.8& 0.619&
    2&84& 0.899& 1.7 \\
-- & 200&  6~~ & 25& 80& 1& 0& 1.7& 0.600& 52&89& $4\times10^{-9}$& &
    1.7& 0.640& 20&03& 0.0055& 1.0 \\
-- & 200& 30~~ & 60& 80& $-1$& 0& 2.1& 0.615& 9&84& 0.198& & 2.1& 0.654&
    15&46& 0.0305& 2.3 \\
-- & 200&  6~~ & 25& 80& $-1$& 0& 1.7& 0.600& 52&89& $4\times10^{-9}$& &
    1.7& 0.640& 20&03& 0.0055& 2.3 \\
 \hline
 \end{tabular}\par}
$N_s$ is the number of spots in the model; ${\langle}R_{\min},R_{\rm
max}{\rangle}$ is the range of radii; $\theta_0$ is observer
inclination; $\alpha_n,$ $\alpha_i$ are parameters specifying
distribution of the spots (eq.~[\protect\ref{alphan}]); $\alpha$ and
$\tilde{\alpha}$ are the power-spectrum indexes (eqs.~[\protect\ref{alpha}]
and [\protect\ref{zb}]); $\sigma$ (eq.~[\protect\ref{sigma}]), $\chi^2$ and
$p$ (eq.~[\protect\ref{chi2}]) characterize the phase properties of the
model.
\end{table*}

\begin{description}
\item (a)~Light curve in the time domain. The curve is mean subtracted
and normalized to the maximum counting rate. A detail of the curve which
covers several periods is shown. Naturally, the light curve of a
single spot is strictly periodic.
\item (b)~Power spectrum $P(\omega)$ in the log-log plane.
Frequency range is chosen to cover $\langle\omega_{\rm
min},\omega_{\max}\rangle$
(a small overlap is necessary because, with only one spot, boundary
frequencies degenerate to $\omega_{\min}=\omega_{\max}$ and
$R_s=R_{\min}=R_{\max}$).
The main peak in the power spectrum corresponds
to the fundamental frequency while the smaller peaks are harmonics.
\item (c)~Fourier transform in the complex plane. While this graph
often exhibits interesting patterns, it is difficult to interpret.
Circles correspond to frequency points
$\omega\leq\omega_{\max}$; crosses correspond to higher
frequencies (they are concentrated with a small amplitude near the origin).
\item (d)~Histogram of the Fourier phases, $\phi(\omega).$ The graph
is normalized to unity and shows a relative distribution which
can be compared with other graphs representing different numbers
of the phase points. Unequal distribution indicates that the phase points
form clusters around certain frequencies. On the other hand, phase
incoherence yields a uniform distribution of the histograms. Frequency interval
$\langle\omega_{\min},\omega_{\max}\rangle$ is again considered.
Recall that points $\phi\pm180\dg$ are to be identified.
\item (e)~Cumulative distribution of the Fourier phases. This graph
contains the same information as the graph (d), but it can be directly
compared with Fig.~4 of Krolik \etal{} (1993). A significant deviation
of the cumulative distribution from the diagonal line (dashed)
indicates the phase coherence and can be formally tested by the
Kolmogorov-Smirnov statistic.
\item (f)~Graph of individual phase points as a function of frequency.
Phase coherence is visible in this graph where the points form
several evident sequences.
\end{description}

Fig.~\ref{fig1} shows a pure light curve (no additional noise). In order
to simulate a high-frequency noise, either inherent to the signal or due
to instrumental errors, we also superposed simulated data with the noise
(with zero mean and a small, typically 5$\,$\% standard deviation).
Typical situations are illustrated in Figures \ref{fig2}--\ref{fig3}
(where the phase coherence is present) and Fig.~\ref{fig4} (where no
phase coherence is indicated). Parameters of these figures are given in
Table~\protect\ref{tab1}.

The first row of Tab.~\ref{tab1} corresponds to
Fig.~\ref{fig1} while subsequent rows describe situations with
much higher numbers of spots which, together with a noisy contribution,
form the predicted fluctuating signal of the source.
The power-spectrum index, $\alpha$ (eq.~[\ref{alpha}]), is also calculated.
The value of $\alpha$ (the slope of the power spectrum as predicted by
the current work) can be compared with a corresponding quantity
$\tilde{\alpha}$ in the approximation of Zhang \& Bao (1991):
\beq
 \tilde{\alpha}=\sfrac{2}{3}\left(4-2\alpha_i-\alpha_n\right)-1,
 \label{zb}
\eeq
which is obtained from their eq.~(3.14) by setting $\alpha_\Omega=3/2$
and $\alpha_t=0$. Eq.~(\ref{zb}) for $\tilde{\alpha}$ has been derived
{\it analytically\/} but with a number of simplifying assumptions about
the light-curve profiles. On the other hand, corresponding values of the
power-spectrum slope $\alpha$ follow from our {\it numerically\/}
derived fitting-formula (\ref{fit}).  Let
us note that $\tilde{\alpha}$ does not depend on observer inclination
while our $\alpha$ shows this dependence. The mean value of the slope is
typically in the range $1\lta\langle\alpha\rangle\lta4,$ but
$\langle\alpha\rangle\lta2$ whenever the inclination is restricted to
$\theta_0\gta75\dg.$ Power spectra become flatter at higher frequencies.
Power-spectrum index is rather sensitive to the spot distribution as
characterized by $\alpha_n$ $(\alpha_n=0$ for the uniform distribution).
A general trend of $\tilde{\alpha}$ being anti-correlated with
$\alpha_n$ (Eq.~\ref{zb}) remains preserved in our analysis. One needs
to be careful in making direct comparisons between $\alpha$
and $\tilde{\alpha}$ because we always considered very restricted range
of $R_s\lta R_{\max}$ (presumably a reasonable restriction for studying
the X-ray variability). In addition there are further, more subtle
differences betwen previous works and our present approach: the fitting
formula (\ref{fit}), finite sizes of the spots, and a numerical package
we have employed ({\sc{}Matlab} v.~4.0 with the usage of its built-in
procedures) but we do not expect these differences to affect the results
significantly. (Indeed, Tab.~1 only briefly summarizes typical results
of the systematic search over the parameter space; lengthy tabular
material can be found at our Web-site,
``http://otokar.troja.mff.cuni.cz/\linebreak[2]user/\linebreak[2]karas/\linebreak[2]au\_www/\linebreak[2]karas/\linebreak[2]papers.htm''.)
We currently work on a detailed comparative analysis of different
approaches which should determine the range of applicability of the
simplified analytic formula~(\ref{zb}).

\begin{figure}
 \vspace{7.8cm}
\caption{No phase coherence is seen in this figure. This is also indicated by
formal tests: $\sigma,$ $\chi^2,$ and $p$ (cp.\
Tab.~\protect\ref{tab1}). The resulting signal corresponds to a large
number of spots $(N_s=200)$ distributed over
a rather broad range of radii $(6\,R_g\leq R_s\leq25\,R_g)$ plus the
$5\,$\% noise component. Moderate inclination $(\theta_0=20\dg)$ means
that variability due to orbital motion is much reduced and relativistic effects
unimportant. We conclude that for these
reasons the phase coherence has been suppressed.}
\label{fig4}
\end{figure}

\section{Conclusion}
Within the considered range of frequency, $10^{-5}$--$10^{-2}\,$Hz,
observations indicate $1\lta\alpha\lta2$ and, very preliminary, no phase
coherence in the Fourier spectra of the signal from AGNs. We have seen
that the bright-spot model is restricted (but not excluded) by these
facts. It is suggested that the procedure described above can be used to
contrast theoretical models of AGNs variability with observations. {\it
These models are testable.} Our present work was motivated by the
bright-spot model of X-ray featureless variability in its original
formulation with a random distribution of individual spots. One might
speculate that the phase coherence will be more important in a more
specific framework which assumes global spiral shock waves extending
across a range of radii in the disc. Phenomenologically, the latter
model can be considered as a subset of the bright-spot scenario where
the azimuthal position of individual spots coincides with a shock wave
propagating through the disc material.

Different assumptions stand behind the two approaches discussed in this
work, the simplified analytic approach of Abramowicz \etal{} (1991) and our
present numerical treatment. Apparent discrepancies between the
corresponding values of the power-spectrum indexes, $\alpha$ vs.\
$\tilde{\alpha}$, call for deeper investigation.

Our current contribution employed energy-integrated light curves. It
should be noted that light curves resolved in time and energy contain
more information: Standard approaches of the Doppler tomography could be
then used to determine structure on the surface of accretion discs
(Karas \& Kraus 1996). ASCA observations offer such data (e.g. Iwasawa
\etal{} 1996) but the resolution has not been enough yet to carry out
this analysis.

The author acknowledges helpful suggestions and critical remarks from
professor A.~Lawrence and from participants of the {\it Trieste-Nordic
Workshop on the Bright-Spot Model 1996}, in particular professors
M.~Abramowicz, G.~Bao, A.~Lanza, and P.~Wiita. This work has been
supported by the grants GACR 205/\linebreak[2]94/\linebreak[2]0504 and
GACR 202/\linebreak[2]96/\linebreak[2]0206 in the Czech Republic.


\label{lastpage}
\end{document}